# Giant Second Harmonic Generation from 3R-MoS$_2$ Metasurfaces

Yilin Tang[1,2], Hao Qin[1], Domenico de Ceglia[3,4,5], Wenkai Yang[1], Mohammad Ali Shameli[6], Mudassar Nauman[1], Rocio Camacho Morales[7], Jingshi Yan[7], Chuanyu Wang[1], Shuyao Qiu[1], Jiri Janousek[1,2], Dragomir Neshev[7*] and Yuerui Lu[1,2*]

[1]School of Engineering, College of Engineering, Computing and Cybernetics, the Australian National University, Canberra, ACT, 2601, Australia

[2]ARC Centre for Quantum Computation and Communication Technology, the Australian National University, Canberra, ACT, 2601, Australia

[3]Department of Information Engineering - University of Brescia, Via Branze 38, Brescia, 25123, Italy

[4]Istituto Nazionale di Ottica - Consiglio Nazionale delle Ricerche, Via Branze 45, Brescia, 25123, Italy

[5]Consorzio Nazionale Interuniversitario per le Telecomunicazioni (CNIT), Viale G.P. Usberti 181/A Sede Scientifica di Ingegneria-Palazzina 3, 43124 Parma, Italy

[6]Faculty of Electrical Engineering, K.N. Toosi University of Technology, Tehran, 16317-14191, Iran

[7]ARC Centre of Excellence for Transformative Meta-Optical Systems (TMOS), Research School of Physics, The Australian National University, ACT 2601, Canberra, Australia

* To whom correspondence should be addressed: Yuerui Lu (yuerui.lu@anu.edu.au) and Dragomir Neshev (dragomir.neshev@anu.edu.au)

**Abstract: Metasurfaces have long served as a cornerstone technique to enhance nonlinear processes, enabling frequency conversion, efficient light manipulation and integrated photonic devices. However, traditional bulk materials often suffer from high absorption losses, hindering the second harmonic generation (SHG) efficiency. Here, we develop a novel approach exploiting quasi-bound state in continuum (qBIC) to achieve giant SHG efficiency in metasurfaces utilizing 3R-MoS$_2$, with high index, superior damage threshold and inherent nonlinearity. The high refractive index of 3R-MoS$_2$, facilitates the high-quality factor (Q) metasurfaces, leading to reduced radiation leakage and localized light confinement within qBIC resonances, with which a remarkable 2000-fold enhancement in SHG intensity has been experimentally demonstrated. Additionally, the twist angle between the lattice orientation and the metasurface unit geometry exhibits a 120°**

**periodicity in its influence on SHG behaviour. By strategically designing to realize the qBIC and exciton dual resonances and optimized twist angle (30°), SHG conversion efficiency was boosted to ~1%, which is around 2 orders of magnitude higher than those of the best metasurfaces on traditional bulk materials. This approach enables potential applications in various areas of nonlinear optics, including frequency conversion, light manipulation, integrated photonics, and quantum communications.**

## Introduction

The field of nonlinear optics is revolutionizing technologies like laser frequency doubling, nonlinear imaging, and quantum information processing.[1,2] However, traditional approaches struggle to achieve high conversion efficiency, limiting their practical applications.[3] Recently, scientists developed a new strategy to overcome this challenge by leveraging the unique properties of quasi-bound states in the continuum (qBIC).[4,5] qBIC offers a tantalizing platform for light-matter interactions, empowering exceptional control over nonlinear processes within meticulously designed metasurfaces.[6,7] Their superior potential arises from the ability to tailor the local electromagnetic field at the nanoscale, effectively integrating with the light-on-chip.[8] The synergistic combination of these two platforms leverages the nanoscale field confinement facilitated by metasurfaces and the large quality factor of qBICs resonances, ultimately leading to drastically enhanced nonlinear optical responses.[9-12] By enabling strong light confinement within the metasurfaces, the linear transmission behaviour with high-quality factor (Q) design directly translates to excellent SHG performance and high conversion efficiency.[11,13-16]

Conventional III-V materials, commonly used for nonlinear applications, suffer from strong absorption in the visible and near-infrared range, particularly at the second harmonic (SH) wave generated *via* SHG.[17,18] This absorption significantly reduces the overall conversion efficiency. Lithium Niobate (LN) is another widely used material known for its strong nonlinearity. However, LN has a relatively modest quadratic nonlinear coefficient, hindering SHG's conversion efficiency.[19,20] Beyond conventional bulky structures, two-dimensional (2D) materials have emerged as promising candidates due to their inherent nonlinearity and susceptibility.[21-24] One key advantage of transition metal dichalcogenides (TMDs) is their high refractive index and second-order susceptibility, enabling the design of high Q metasurfaces. The work of Nauman et al. demonstrated that these properties coupled to a Mie-resonant TMDc metasurface allow for highly directional nonlinear light emission at the nanoscale.[25] The high refractive index of TMDs strengthens their potential for high-quality resonances and efficient SHG applications within metasurfaces due to the significantly enhanced electromagnetic field confinement.[26] Furthermore, TMDs offer several merits over conventional materials. They exhibit a higher damage threshold, allowing for higher power operation without material degradation. Unlike III-V materials, which only host excitons at cryogenic temperatures,[27,28] TMDs exhibits tightly bound excitons at room temperature,[29-31] which can be used to enhance SHG.[32] While conventional 2H-$MoS_2$ exhibit diminished nonlinearity due to restored centrosymmetry[34], the "AA" stacking of the 3R polytype of molybdenum disulfide (3R-$MoS_2$), preserves the broken inversion symmetry, allowing the atomic nonlinear dipole to constructively build up layer by layer, leading to enhanced total nonlinear gain.[35-42] This inherent nonlinearity makes 3R-$MoS_2$ an ideal candidate materials for the qBIC metasurfaces.

In essence, this work weaved together the remarkable nonlinearity of 3R-$MoS_2$, the dual resonances of high-Q qBICs and exciton, and the tailored nonlinear susceptibility provided with twist angle between metasurface structure and lattice orientation of 3R-$MoS_2$, culminating an unprecedented enhancement of SHG. We also delved deeper into the influence of the twist angle, revealing the interplay between interlayer electronic coupling and the nonlinear response, highlighting the importance of considering both geometric design and lattice orientation in optimizing SHG efficiency. With strategic optimization of the twist angle and the dual resonances of the exciton and qBIC, our metasurface I showed an unprecedented SHG conversion efficiency ~1%, surpassing previously reported best metasurface by around two orders of magnitude.[25,43-51] This landmark accomplishment represents a significant leap forward, pushing the boundaries of metasurface-based SHG and unlocking the transformative potential for future efficient nonlinear optical devices, from ultrafast signal processing to advanced bioimaging and light manipulation at the nanoscale.[52-54] The insights gleaned from this work illuminate the path towards further optimization of metasurface-based SHG platforms and the development of next-generation devices with exceptional capabilities in high-harmonic generation, laser frequency doubling, nonlinear imaging, and quantum information processing.[55-59]

## Results and Discussion

**Concept of Quasi-BIC 3R-MoS$_2$ Metasurface**.

The choice of materials for the metasurface elements plays a significant role on performance of quasi-BIC metasurfaces. Low-loss and high-index materials contribute to a longer lifetime of the quasi-BIC and minimize losses in order to achieve a high-$Q$ factor or efficient nonlinear interactions.[5,60-62] 3R-MoS$_2$, a 2D material exhibiting strong non-linearity, high-index and superior damage threshold in near infrared region, is quite a favourable material for achieving tailored optical responses suitable for quasi-BIC design.[36] Moreover, the atomic structure of the 3R-MoS$_2$ produces the same multiplicative dipole orientation in each layer, ensuring an enhanced stacking effect between stacking layers, leading to a progressively stronger SHG signal with increasing layer numbers (Fig. 1a inset). We have engineered the metasurface to use a repeating pattern of crescent-shaped resonators (Fig. 1a and 1b). By varying the thickness $t$ of 3R-MoS$_2$ flake as well as the resonator height, the radius $r$ of the crescent, and the periodicity $P_x = P_y$, we can precisely tune the qBIC resonance (Fig. 1b). Our fabricated 3R-MoS$_2$ metasurface A exhibits a sharp transmission dip centred at 1655 nm in the measured linear transmission spectrum, coinciding with one of the resonances of the structure (Fig. 1c), when the pump incident is polarized along y axis of the crescent unit geometry, corresponding to $\theta = 0°$ ($\theta$ is the angle between the polarization of pump incident and y axis of metasurface geometry.) (Fig. 1c inset).[15] The qBIC mode showed a measured $Q$ value of 162, which suggests a high confinement of light within the metasurface with strong field enhancement. The simulated transmission spectrum with a resonance peak at 1655 nm matches reasonably well with the measured qBIC resonance (Fig. 1d and Supplementary Note 3). The simulated distribution of the magnetic and electric fields (Fig.1e and f) shows that the qBIC associated to a magnetic dipole mode is a primary driver of qBIC resonance in our 3R-MoS$_2$ metasurface A. Due to the presence of a strong z-component of the magnetic resonance and the circulating electric field, we conclude that the resonance is mainly associated with a vertical (z-oriented) magnetic dipole. This mode would be dark in a symmetric resonator under plane wave illumination at normal incidence (symmetry protection), while it turns bright in the crescent nanoparticle due to the symmetry breaking. The spectral signature associated with the excitation of the mode is a Fano resonance (qBIC) with its typical asymmetric spectral shape.[63-66]

**Enhanced SHG performance of 3R-MoS$_2$ metasurfaces**

The crescent-structured 3R-MoS$_2$ metasurface meticulously designed with qBIC resonance serves as a potent platform to precisely control and enhance the SHG process. The linear polarized pump was incident propagated from the back of the sample, through the sapphire substrate before reaching the metasurface (Fig.2a). The experimental results of the measured wavelength-dependent SHG of the metasurface A (qBIC at 1655 nm) depict a significant increasing trend as the incident pump wavelength approaches to the qBIC resonance at 1655 nm (Fig. 2b). The measured power dependent SHG, with a measured fitting slope of ~1.95, confirms the quadratic nature of the SHG process (Fig.S2b). The SHG process is inherently polarization-dependent, and the signature of a qBIC resonance in polarization-resolved SHG would likely involve a distinct dependence on the polarization states of the pump incident (Fig. 2c). The SHG intensity was maximized when the pump incident was polarized along y-axis ($\theta = 0°$) of the crescent, aligning with the dominant resonance mode in transmission spectrum in Fig. 1c. This confirms that the enhanced SHG intensity at particular polarization arises primarily from the symmetry broken induced by the metasurface design. Impressively, when the incident pump wavelength moves close to qBIC resonance of 1655 nm and polarized at $\theta = 0°$, the metasurface has ~330-fold SHG intensity enhancement (in forward direction) compared to its bulk 3R-MoS$_2$ counter partner, even though the bulk 3R-MoS$_2$ flake has ~5.83 times more materials than the nano-structured metasurface. To quantitatively assess the SHG enhancement from the qBIC, we define the enhancement factor as the ratio between the qBIC on resonance and the off-resonance (around 1540 nm) SHG intensities. This definition has considered the small wavelength-dependent second-order nonlinearity $\chi^{(2)}$ of 3R-MoS$_2$ in this wavelength range.[67] The enhancement factor progressively increases as the pump wavelength approaches the qBIC resonance, reaching a peak value of approximately 2088 (Fig. 2d). This enhancement factor is roughly consistent with the number obtained by multiplying the aforementioned ~330-fold enhancement (SHG intensity enhancement from the metasurface) with the material volume ratio factor of 5.83 between metasurface and its bulk 3R-MoS$_2$ counter partner. This observation underscores the robust nonlinear performance of our metasurface and its effectiveness in harnessing the qBIC resonance to dramatically enhance

SHG. We also employed a rigorous numerical approach to simulate the SHG process in 3R-MoS$_2$ metasurfaces, accounting for the nonlinear polarization induced by the fundamental field and the specific nonlinear susceptibility tensor of 3R-MoS$_2$ (Supplementary Note 4).

The resonance mode could be strategically tuned to a particular wavelength position by manipulating periodicity, and the symmetry of the resonators at a particular wavelength position of the 3R-MoS$_2$. We have conducted a comprehensive parametric study to investigate the influence of geometric parameters, such as displacement, periodicity, thickness, and radius, on the qBIC resonance wavelength (Fig. S3 and Supplementary Note 3). Precise control over these parameters allows for engineering the resonance and tailoring the optical response. We have also fabricated a few more 3R-MoS$_2$ metasurfaces (B, C, D) with qBIC mode measured at 1514 nm, 1600 nm, and 1648 nm, respectively. Their SHG enhancement factors are 1684, 1100, and 1798, respectively (Fig. S4).

The $Q$ values of our 3R-MoS$_2$ metasurfaces were designed and fabricated to be 100-150, matching well with the femtosecond laser bandwidth in our measurements. The $Q$-factor of the qBIC can be designed to be a few orders of magnitude higher if needed, enabling low-power nonlinear devices based on 3R-MoS$_2$ metasurfaces (Fig. S1g). Our numerical simulation results indicate that the main handle of control is the asymmetry of the resonator, which in our structure is the displacement (Fig. S3a and Supplementary Note 2). We also conducted numerical simulations to investigate the impact of geometric imperfections on the performance of 3R-MoS$_2$ metasurfaces. Our results indicate that the resonance wavelength associated with the qBIC is not severely affected by roughness and slant of the side walls (Fig. S1 and Supplementary Note 2). However, the roughness of the sidewall (Fig. 1b), one of the common fabrication imperfections, introduces asymmetry that in turn tends to lower and limit the quality factor of qBICs. For a displacement equal to 286 nm, as in the fabricated samples (Fig.1b), the simulated $Q$ decreases from a value of 680 in the smooth case to a value of 250 in the rough case. Since the SHG efficiency scales as the square of the quality factor of the resonator, the effect of roughness causes a decrease in SHG efficiency by a factor slightly larger than 7. We can further improve the etching recipes to increase the $Q$ factor and thus SHG.

**SHG Modulation from 3R-MoS$_2$ metasurfaces induced by the twist angle $\varphi$**

In order to systematically investigate the influence of the crystalline lattice orientation on the enhanced nonlinear performance of qBIC metasurface, we define script phi ($\varphi$) as the twist angle between the lattice armchair direction and the *y*-axis of the metasurface unit cell (Fig. 3a inset). In order to exclude potentially confounding factors, we precisely fabricated the two metasurface arrays with identical design patterns on the same flake, both showing the resonances closed to 1330 nm but with different twist angles, $\varphi = 30°$ and $\varphi = 60°$, respectively. Any minor discrepancies (such as variation in flake thickness, crystal defect, nonuniform etching, surface roughness, etc) might be induced by fabrication imperfection. A simulated SHG intensity mapping as a function of $\varphi$ and $\theta$ was conducted to gain a comprehensive understanding on the influence of the crystalline orientation on the qBIC resonance (Fig. 3b). Notably, with $\theta = 0°$, the SHG intensity exhibits a periodic behaviour with a period of 120° and peaks at specific $\varphi$, with the maximum occurring at the $\varphi = 30°$ (equivalent to 150°, 270°…etc).

Quantitative analysis reveals a striking enhancement in the qBIC-induced SHG for these distinct twist angles. Compared to $\varphi = 60°$, the SHG at $\varphi = 30°$ has an overall 4.5 times increase (Fig. 3c-d), consistent with the simulation (Fig. 3b). This highlights the crucial role of $\varphi$ in modulating the SHG performance. Under this certain scenario, all other relevant fabrication parameters and material properties are rigorously controlled and remain the constant, highlighting the twist angle between the metasurface and the underlying 3R-MoS$_2$ lattice as the key determinant of the observed SHG intensity disparity. We have further investigated the underlying physics by analyzing the eigenmodes. The strong dependence of the SHG efficiency on the twist angle $\varphi$ can be explained by the orientation of the nonlinear polarization relative to the eigenmodes of the metasurface (Fig. S7 and Supplementary Note 4-5).

**Highly enhanced SHG from 3R-MoS$_2$ metasurface with dual resonances of qBIC and exciton**

Exciton resonance in TMD semiconductors can be used to increase the SHG efficiency.[32] The SHG efficiency of 3R-MoS$_2$ metasurfaces can be further enhanced, with the design of dual resonances of qBIC (with excitation photons) and exciton (with SH photons).[68] Combining with a comprehensive scan of the incident pump wavelengths on 3R-MoS$_2$ bulk flake, the position of exciton peak ($E_0^A$) at room temperature could be estimated around roughly ~1.84 eV (~675 nm), consistent with previous report,[69] which is indicated by a discernible increase SHG intensity when the SH wavelength is approaching to the $E_0^A$ (Fig. S10a, Supplementary Note 6). The design parameters were systematically tuned to attain the resonance close to $E_0^A$ (Fig. 4a), and we have successfully fabricated metasurface F (qBIC at 1372 nm) (Fig. 4b, S1a). The maximized SHG intensity was observed when the incident pump approaches to the resonance, demonstrating superior absolute intensity compared to other metasurfaces (Fig. 4c-d, Fig. S8). There was actually still a slight gap between the $E_0^A$ of 3R-MoS$_2$ (~675 nm) and half of wavelength of qBIC resonance (~686 nm) in metasurface F. Due to the large binding energy of excitons in TMDCs, the excitons can exist at room temperature or even higher temperature, in contrast to III-V semiconductors. This allows dynamic tuning of exciton resonance with temperature over a wide range, facilitating precise alignment of those two resonances. Notably, the exciton peak in 3R-MoS$_2$ exhibits significant temperature dependence.[70,71] On the other hand, temperature variations (from 20°C to 200°C) have minimal impact on the spectral position of the qBIC, due to the low thermo-optic coefficient of 3R-MoS$_2$ (Fig. S9). This property allows for independent manipulation of the SHG process. In other words, we can spectrally tune the SHG signal, by tuning the exciton resonance as a function of temperature, without affecting the qBIC resonance. Therefore, we have conducted SHG measurement as a function of temperature, aiming at further boosting the SHG intensity and conversion efficiency by tuning the $E_0^A$ position. When the temperature increased from room temperature to 100°C, the exciton peak of 3R MoS$_2$ was shifted from ~675 nm to ~686 nm (Supplementary Note 6), making the exciton and qBIC dual resonances well matched at 100°C, leading to a significant enhancement in SHG efficiency by approximately threefold comparing with that at room temperature (Fig. 4c).[69] This significant SHG enhancement is caused dominantly by the temperature-induced exciton shift, because the second order nonlinearity of the 3R-MoS$_2$ at off-resonance bands showed a small temperature dependence (Fig. S10b).

Achieving high absolute SHG conversion efficiency has remained a significant challenge in nanoscale devices that do not rely on phase-matching techniques. Our approach demonstrates a breakthrough in this area. Conversion efficiencies were measured and rigorously compared across various control groups, where monolayer under 1550 nm pump only exhibited a conversion efficiency of $10^{-9}$ with a power density of 33.56 $GW/cm^2$ (Fig. 4f) which is less than the damage threshold. Notably, under the same power density, the conversion efficiency from 3R-MoS$_2$ bulk flake as the control group of metasurface A reached $10^{-7}$ while that from the metasurface A (qBIC at 1655 nm) has relatively higher conversion efficiency of $10^{-5}$, albeit limited due to the resonance being distant from the $E_0^A$. Remarkably, the conversion efficiency from the metasurface F is approaching 0.1% measured at room temperature. Accordingly, at 100°C the estimated conversion efficiency surged to over 0.2%. With the exciton and qBIC dual resonances and a twist angle $\varphi$ of 30°, our metasurface I (Supplementary Note 7) has achieved a conversion efficiency approximately 1% (Fig. 4f and Fig. S11b), surpassing previously reported best metasurface by around 2 orders of magnitude (Fig. 4f and Table S1).[44-46,48,49,51,72-74] This underscores the potential for enhancing SHG performance through the mode matching of qBIC resonance and $E_0^A$. This strategy has the potential to achieve improvements on developing highly efficient, tunable SHG devices. The dual resonance (qBIC + exciton resonance) might result in limited wavelength choices for practical applications, as the exciton position of a specific material is normally fixed. However, we could use 3R phase TMD alloys, for example, 3R-Mo$_{(1-x)}$W$_{2x}$S$_2$ ($x$ = 0-1), which offer tunable and broad range of exciton energy positions as a function of alloy composition $x$, allowing us to realize a response tailored to specific applications.[33]

**Conclusion**

This work demonstrates the remarkable potential of qBIC metasurfaces for achieving significant enhancements in SHG efficiency. This exceptional performance stems from the synergistic effect of high index of 3R-MoS$_2$, superior damage threshold, inherent exciton resonance at room temperature, and engineering capability of lattice orientation. We have experimentally demonstrated a 2000-fold increase in SHG intensity at the qBIC resonance in meticulously designed 3R-MoS$_2$ metasurfaces. The critical role of the twist angle between the metasurface structure and lattice orientation of 3R-MoS$_2$ in modulating SHG performance has been further unveiled. This twist angle modulates SHG intensity by a factor of 4-5, offering a powerful parameter for performance optimization. Finally, we have optimized the design to strategically engineer the qBIC resonance and thermally tune the $E_0^A$, realizing dual resonance and consequently boosting the SHG performance and conversion efficiency exceeding 0.2%.

Strategic engineering of the qBIC resonance and twist angle has boosted the conversion efficiency approaching 1%, surpassing previously reported best metasurface by around 2 orders of magnitude (Fig. 4f and Table S1). These findings enable further exploration of qBIC metasurfaces for efficient nonlinear light manipulation and advanced photonic devices.[75,76]

**Materials and Methods**

**Numerical simulation**

In the linear simulation, we used the full-wave software of COMSOL Multiphysics. We excited the plane wave with linear polarization along the y direction. The resonators have a thickness of 213 nm and the periodicity is 760 nm; the radius and displacement of the moon-shaped resonators are 260 nm and 286 nm, and the thickness of the substrate is 1500 nm. The boundary conditions are set as periodic; thus, the structure is supposed to be infinite in the direction of the x- and y-axis. Using tetrahedral meshes and investigation based on finite element method (FEM) the transmission of the metasurface is calculated and depicted in Fig. 1. As shown in this figure, the numerical results based on the linear simulation and the experimental data are close to each other.

**Metasurface fabrication**

The metasurfaces were fabricated on Sapphire substrate. The 3R-$MoS_2$ flakes were mechanically exfoliated to ensure a higher quality and then transferred onto the pretreated substrates. The flake thicknessed were measured through the surface profiler. Next, the ZEP positive resist (ZEP 520A) was spin-coated onto the substrate and followed by a thermally deposited 10nm gold layer as a conductive layer. The patterns were then defined by electron-beam lithography (EBL). A 40 nm thick Al layer was subsequently deposited by e-beam evaporation after developing the sample using the developer (ZED N50- (n Amyl Acetate)). Following that is the lift-off process during which samples are soaked in a resist remover

(ZDMAC- (Di-Methyl Acetamide)). The remaining Al as the etching mask in the subsequent inductively coupled plasma (ICP) etching. The Al cap was then removed by the Al etchant. The topography/morphology of the metasurfaces were characterized through scanning electron microscopy (SEM).

**Linear transmission measurement**

The polarization-resolved transmittance spectra were collected by a spectrometer (Ocean Optics QE65000) equipped with white light source. A broadband polarizer was set after the light source to select the polarization of the pump incident light to excite the qBIC resonance. A convex lens was used to focus the incident light to the sample plane with normal (0°) incidence and a 20× objective lens after the samples was used to collect the transmitted light. The transmittance spectra were normalized with respect to the transmittance through the substrate. The size of white light spot in our setup can be tuned from 10 to 100 µm. In our measurements, we tuned the spot size of the white light source to be around the array size of the metasurface (Table S2).

**SHG measurement**

The wavelength-dependent SHG was measured with a pulsed laser (Spectra-Physics Mai Tai with repetition frequency: 80 MHz, pulse duration: ~200 fs) operating with an OPO to output the tunable fundamental pump wavelength. The spectral of the pump laser ranges from 1300 to 1700 nm. The samples were excited *via* a ×50 near infrared (NIR) infinity corrected objective lens (NA = 0.42) and the signal was collected via a ×100 near infrared (NIR) infinity corrected objective lens (NA = 0.5) in the transmission mode. A broadband $\lambda/2$ waveplate was set after the pump incident to align the polarization of the beam with the metasurface unit geometry qBIC design.

**Acknowledgements**

The authors acknowledge funding support from the ANU Ph.D. student scholarship, Australian Research Council (grant No: DP240101011, DP220102219, LE200100032), ARC Centre of Excellence in Quantum Computation and Communication Technology (project number CE170100012), and the National Health and Medical Research Council (NHMRC; ID: GA275784). Ministero dell'Istruzione e del Merito (METEOR, PRIN-2020 2020EY2LJT_002); H2020 Future and Emerging Technologies (FETOPEN-2018-2020 899673, METAFAST); NATO (SPS G5984); European Union under the Italian National Recovery and Resilience Plan (NRRP), project SMART, CUP E63C22002040007, partnership on "Telecommunications of the Future" (PE00000001 - program "RESTART").

**Competing interests**

The authors declare that they have no competing financial interests or any other conflict of interest.

**Author Contributions**

Y. L. conceived the project; Y. L. and D. N. supervised the project; Y. T. fabricated 3R-$MoS_2$ metasurfaces; Y. T., H. Q., R.M., J. Y., C. W. and S. Q. carried out the optical measurements; Y. T. and Y. L. analyzed the data; H.Q. and J. J. built up the SHG optical setup; D. d. C., M. A., M. N. and W. Y. run the simulation; Y. T. and Y. L. drafted the manuscript and all authors contributed to the manuscript.

**Supplementary Information**

All additional data and supporting information and methods are presented in the supporting information file available online. The figures and information in the supporting information have been cited at appropriate places in this manuscript.

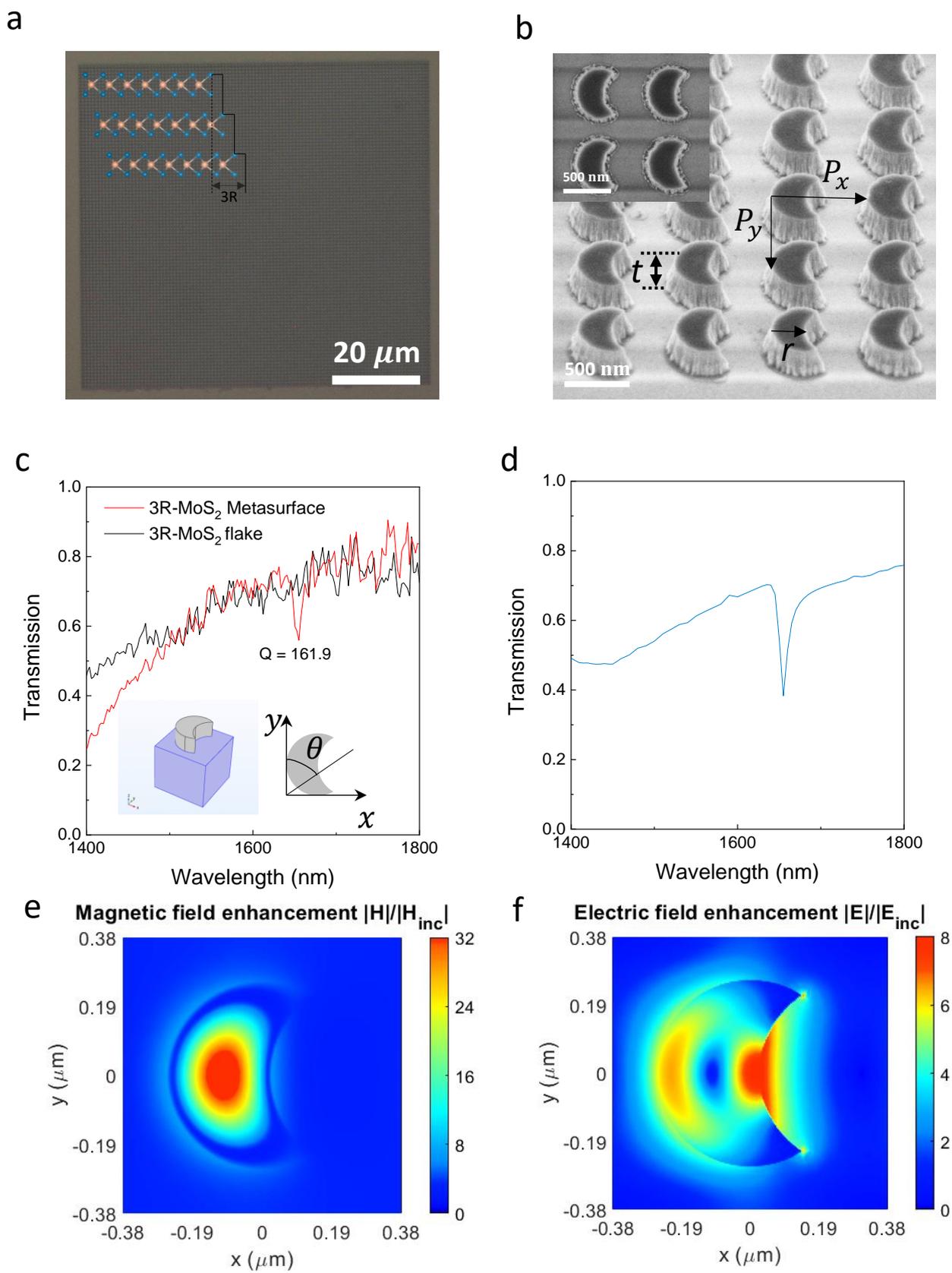

Figure 1

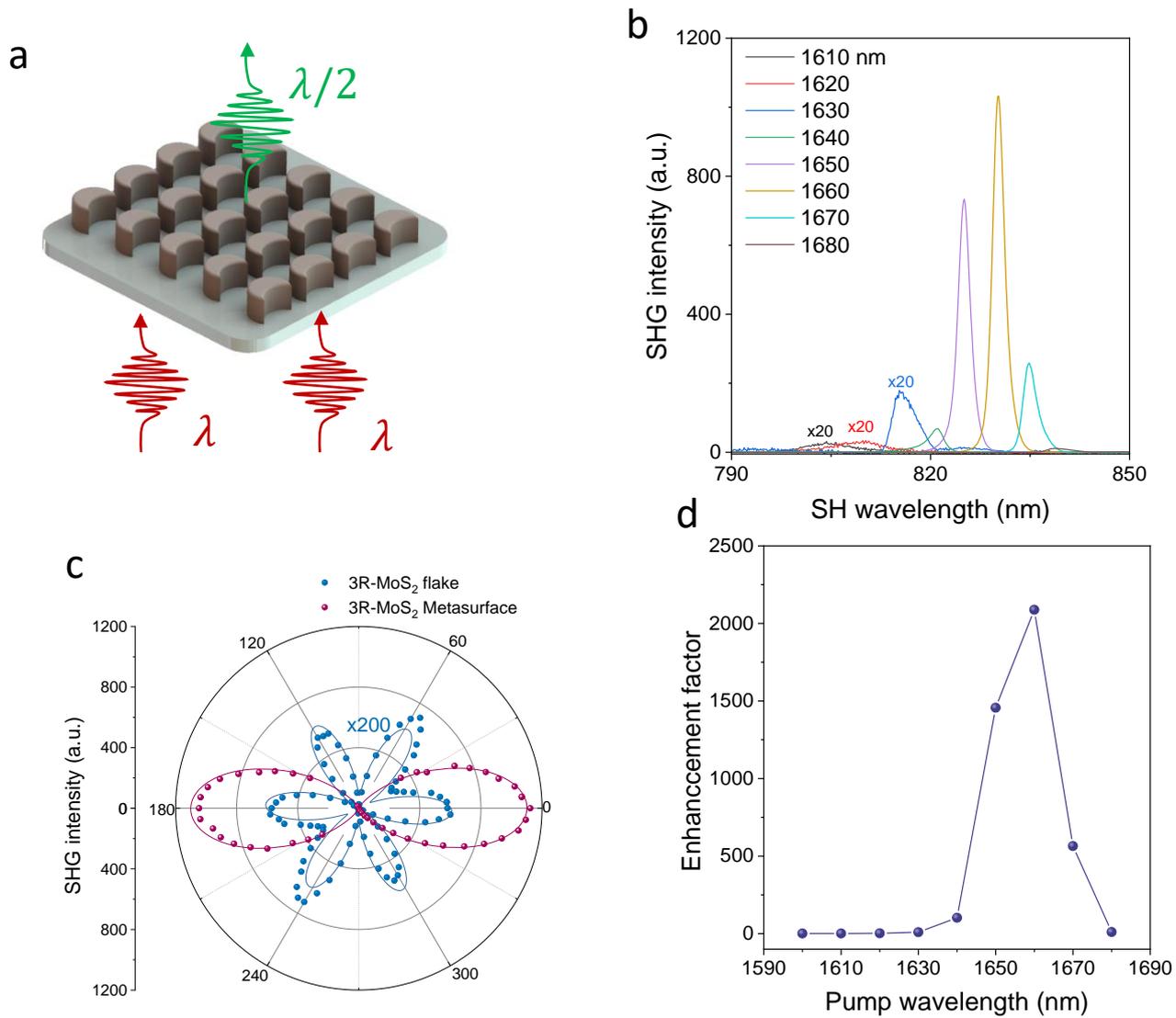

Figure 2

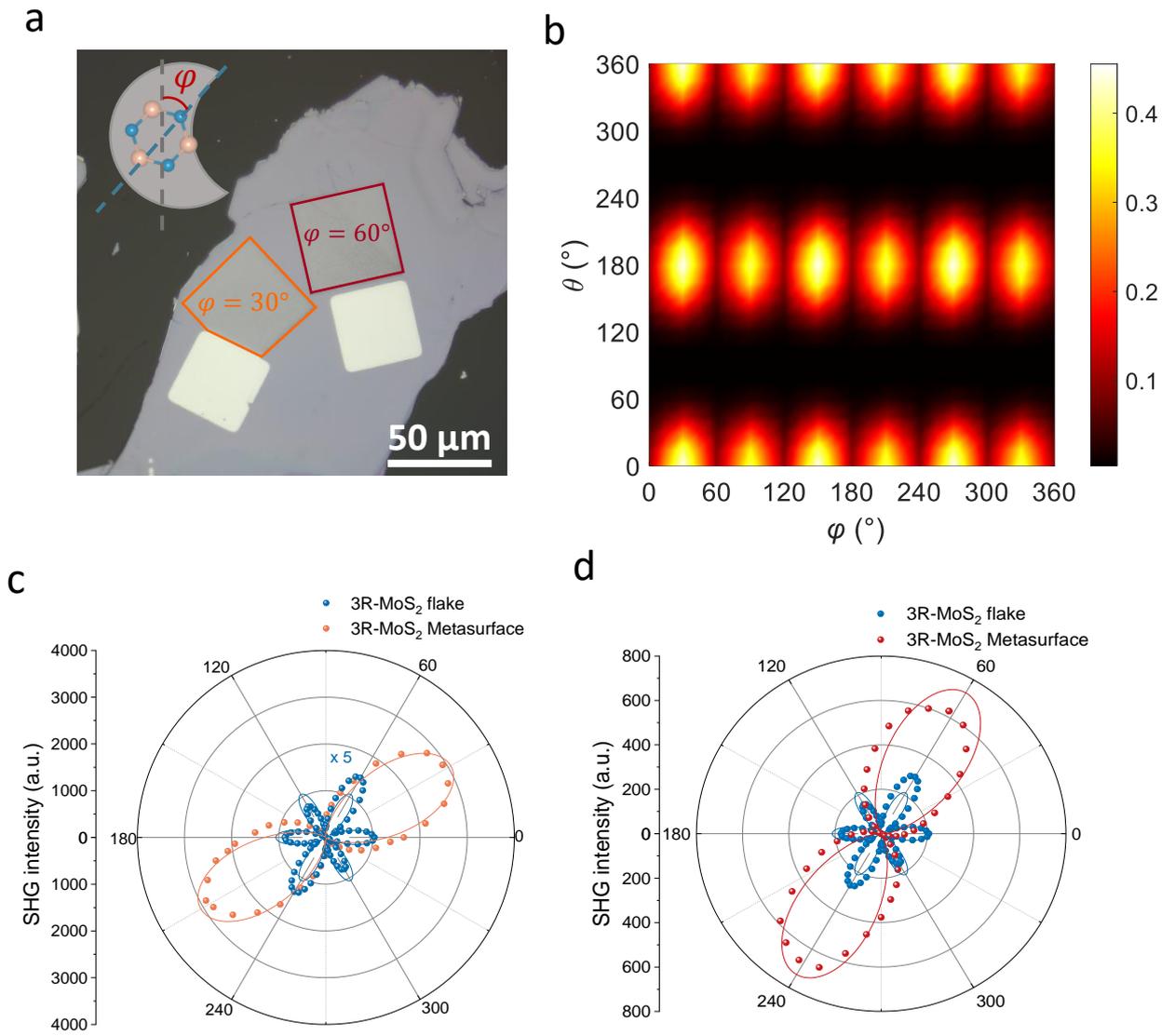

Figure 3

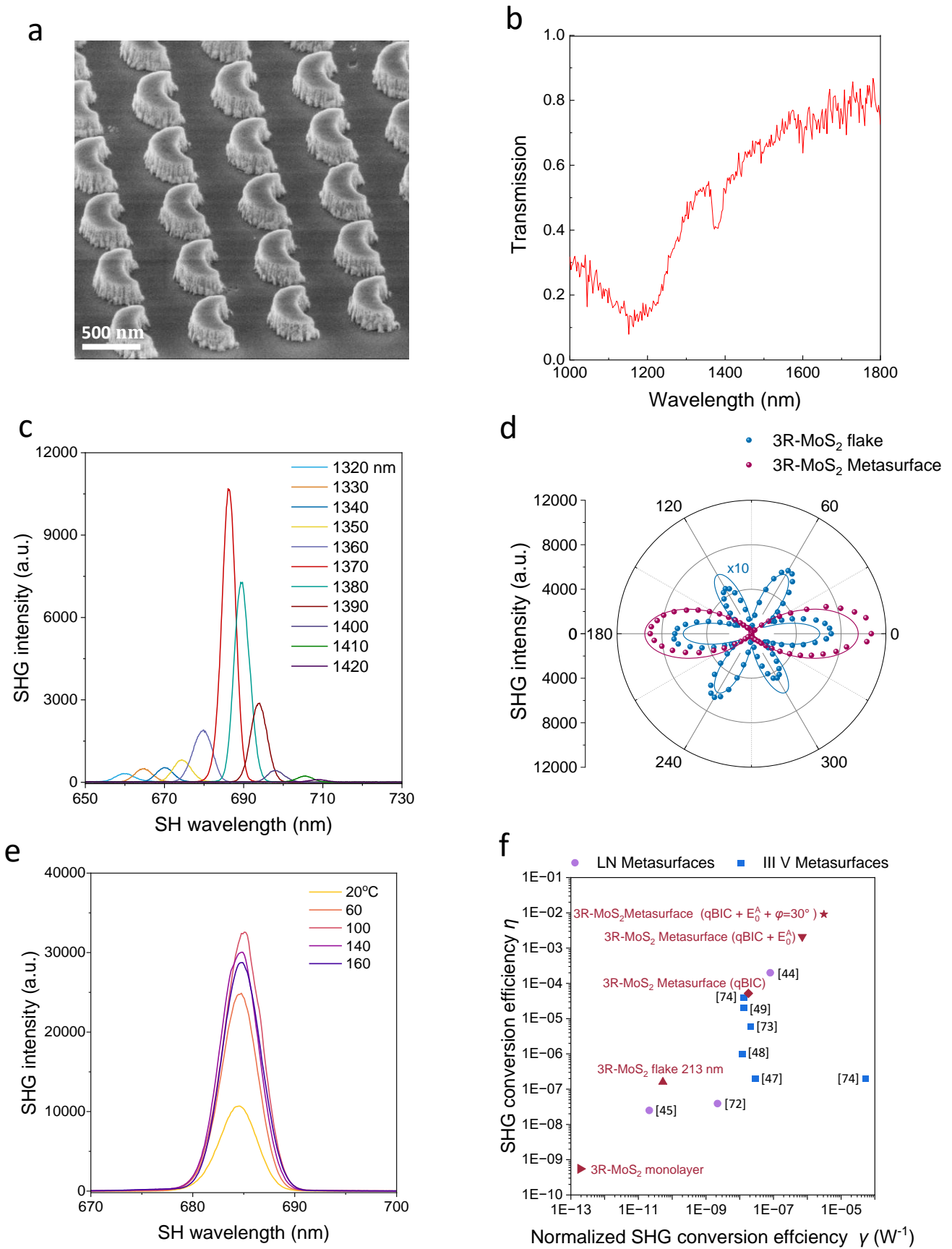

Figure 4

**FIGURE CAPTIONS**

**Figure 1 | Concept of Quasi-BIC 3R-MoS$_2$ Metasurface. a,** Optical image of 3R-MoS$_2$ qBIC metasurface A. Inset is the atomic structure of 3R-MoS$_2$. **b,** SEM image of 3R-MoS$_2$ metasurface. Inset is the top view, where $t$ is the thickness, $r$ is radius, $P_x = P_y$ is the periodicity. **c,** Measured linear transmission spectra of 3R-MoS$_2$ metasurface A (red, qBIC at 1655 nm) and 3R-MoS$_2$ control flake (black), with pump incident light linearly-polarized at $\theta = 0°$. Inset: Metasurface unit. $\theta$ is the angle between the polarization of pump incident and y axis of metasurface geometry. **d,** Simulated linear transmission spectrum of 3R-MoS$_2$ metasurface (qBIC at 1655 nm) at pump incident polarized at $\theta = 0°$. **e-f,** Simulated magnetic (e) and electric field (f) enhancement distribution at the resonant λ of 1655 nm.

**Figure 2 | Enhanced Second Harmonic Generation (SHG) of 3R-MoS$_2$ metasurface A. a,** Schematic illustration of SHG from 3R-MoS$_2$ qBIC metasurface. **b,** Measured SHG intensity as a function of excitation wavelengths from metasurface A, under constant power of 5.9 mW. For better visualization, the intensity of excitation 1600 nm, 1610 nm and 1620 nm was multiplied by a factor of 20. **c,** Measured (dots) and simulated (solid line) co-polarized SHG as a function of the polarization angle of incident laser, from 3R-MoS$_2$ flake (blue), and 3R-MoS$_2$ metasurface A (purple). The SEM image of 3R-MoS$_2$ metasurface A is shown in Fig.1b. In experiment, the polarization of the incident laser, at wavelength of 1655 nm and power of 5.9 mW, was initially set to be along the armchair direction of the flake and controlled by rotating the $\lambda/2$ waveplate. A linear polarizer was used to select the polarization component of the SH radiation parallel to the polarization of the pump beam. **d,** Measured SHG enhancement factor from metasurface A (relative to the average of measured SHG under excitation at off resonance position of ~1540 nm to ensure the data reliability) as a function of pump wavelengths.

**Figure 3 | SHG Modulation from 3R-MoS$_2$ metasurfaces induced by the twist angle $\varphi$. a,** Schematic of the lattice orientation of the bulk 3R-MoS$_2$ with respect to the y axis of metasurface. $\varphi$ is the twist angle between armchair direction of MoS$_2$ lattice and metasurface asymmetric y axis. **b,** Simulated transmitted SHG, as a function of lattice orientation $\varphi$ and pump incident polarized angle $\theta$. **c-d,** Measured polarization dependent SHG, under pumping wavelength of 1330 nm and power of 8.4 mW, of 3R-MoS$_2$ metasurface E (metasurface E$_i$

orange(c) and metasurface $E_{ii}$ red(d)) and 3R-MoS$_2$ flake (blue) as bulk control group with mismatch angle $\varphi = 30°$ (c) and $\varphi = 60°$ (d).

**Figure 4 | Highly enhanced SHG from 3R-MoS$_2$ metasurface with dual resonances of qBIC and exciton. a,** SEM image of 3R-MoS$_2$ metasurface F. **b,** Measured linear transmission spectrum of 3R-MoS$_2$ metasurface F at pump incident along $\theta = 0°$. **c,** Measured SHG intensity as a function of wavelength of metasurface F (qBIC at 1372 nm), under constant power of 7.5 mW. **d,** Measured (dots) and simulated (solid line) co-polarized SHG, under pumping wavelength of 1372 nm and power of 7.5 mW, as a function of the polarization angle of incident laser, from 3R-MoS$_2$ flake (blue), and 3R-MoS$_2$ metasurface F (purple). **e,** Measured SHG intensity as a function of temperature at the pump wavelength of 1372 nm (on resonance) and constant power of 2.5 mW. The SH intensity was maximized at around 100°C. **f**, Comparison between SHG performance of different metasurfaces. Absolute SHG efficiency $\eta$ and normalized efficiency $\gamma$ are shown on the vertical and horizontal axes, respectively. Red markers are 3R-MoS$_2$ monolayer under pump wavelength 1550 nm, 3R-MoS$_2$ flake as bulk control group (213 nm) of Metasurface A under pump wavelength 1655 nm, 3R-MoS$_2$ Metasurface (qBIC) A, Metasurface F (qBIC + $E_0^A$) and Metasurface I (qBIC + $E_0^A$ + $\varphi = 30°$) as labelled at the side of each category. Blue and purple markers are III V Metasurfaces and lithium niobate (LN) Metasurfaces respectively.